\documentclass[11pt]{article}
\usepackage[latin9]{inputenc}
\usepackage{amsfonts}
\usepackage{amsmath}
\usepackage{amssymb}
\usepackage{indentfirst}
\usepackage{graphicx}
\usepackage[colorlinks]{hyperref}
\usepackage{cite}

\numberwithin{equation}{section}
\begin{document}

\begin{titlepage}
\vspace{3cm}
\baselineskip=24pt

\begin{center}
\textbf{\LARGE{On the Maxwell supergravity and flat limit in 2+1 dimensions}}
\par\end{center}{\LARGE \par}

\begin{center}
	\vspace{1cm}
	\textbf{Patrick Concha}$^{\ast}$,
	\textbf{Diego M. Peñafiel}$^{\ast}$,
	\textbf{Evelyn Rodríguez}$^{\dag}$,
	\small
	\\[5mm]
	$^{\ast}$\textit{Instituto
		de Física, Pontificia Universidad Católica de Valparaíso, }\\
	\textit{ Casilla 4059, Valparaiso-Chile.}
	\\[2mm]
	$^{\dag}$\textit{Departamento de Ciencias, Facultad de Artes Liberales,} \\
	\textit{Universidad Adolfo Ibáñez, Viña del Mar-Chile.} \\[5mm]
	\footnotesize
	\texttt{patrick.concha@pucv.cl},
	\texttt{diego.molina.p@pucv.cl},
	\texttt{evelyn.rodriguez@edu.uai.cl},
	\par\end{center}
\vskip 20pt
\begin{abstract}
\noindent
The construction of the three-dimensional Chern-Simons supergravity theory invariant under the minimal Maxwell superalgebra is presented. We obtain a supergravity action without cosmological constant term characterized by three coupling constants. We also show that the Maxwell supergravity presented here appears as a vanishing cosmological constant limit of a minimal AdS-Lorentz supergravity. The flat limit is applied at the level of the superalgebra, Chern-Simons action, supersymmetry transformation laws and field equations.

\end{abstract}
\end{titlepage}
\newpage{}

\section{Introduction}

The three-dimensional (super)gravity is considered as an interesting toy
model for approaching richer (super)gravity models, which in general are non
trivial. Furthermore, the three-dimensional theory shares many interesting
properties with higher-dimensional theories such as black hole solutions
\cite{BTZ}. Remarkably, supergravity theory in three spacetime dimensions
\cite{DK, Deser} can be expressed as a gauge theory using the Chern-Simons
(CS) formalism for the AdS \cite{AT} or Poincaré supergroup \cite{AT2}. In
the last decades, diverse three-dimensional supergravity models have been
studied in \cite{MS, Nieuwenhuizen, RN, NG, HIPT, BTZs, GTW, GS, ABRHST,
ABRS, BKNTM, NT, ABBOS, FMT, FMT2, BRZ, BDMT2, BDR, FMT3}. In particular,
there has been a growing interest to extend AdS\ and Poincaré supergravity
theories to other symmetries.

The Maxwell symmetry has become an interesting alternative for generalizing
Einstein gravity. The Maxwell algebra has been introduced to describe a
particle moving in a four-dimensional background in the presence of a
constant electromagnetic field \cite{BCR, Schrader, GK}. Further
applications of the Maxwell group have been developed in \cite{BH, Sorokas2,
BG}. The Maxwell symmetry and its generalizations have been useful to
recover General Relativity (GR) as a particular limit using the CS and
Born-Infeld (BI) gravity formalism \cite{EHTZ, GRCS, CPRS1, CPRS2, CPRS3}.
In four dimensions, Maxwell gravity models have been successfully
constructed in \cite{AKL, DKGS, AKL2}. More recently, there has been a
growing interest in studying the three-dimensional CS gravity theory
invariant under the Maxwell group \cite{SSV, HR, CCFRS, AFGHZ, CMMRSV}. This
theory, as the Poincaré one, is asymptotically flat and does not contain a
cosmological constant term. Interestingly, a novel asymptotic structure
appears for the three-dimensional Maxwell CS gravity \cite{CMMRSV}. Such
asymptotic symmetry corresponds to a deformed $\mathfrak{bms}_{3}$ algebra
and has been first introduced in \cite{CCRS} by using the semigroup
expansion procedure \cite{Sexp}. Furthermore, it has been shown that the
vacuum energy and angular momentum are influenced by the presence of the
gravitational Maxwell field \cite{CMMRSV}.

The Maxwell algebra also appears as an Inönü-Wigner (IW) contraction \cite%
{IW, WW} of an enlarged algebra known as AdS-Lorentz algebra. The
AdS-Lorentz algebra, also known as the semisimple extended Poincaré algebra,
was introduced in \cite{SS, GKL} and has been studied in the context of
gravity in diverse dimensions \cite{HR, PSSS, DFIMRSV}. Such symmetry can be
written as the semidirect sum $\mathfrak{s0(}d-1,2)\oplus $ $\mathfrak{s0(}%
d-1,1)$ which motivates the name of AdS-Lorentz.

At the supersymmetric level, a minimal extension of the Maxwell algebra
appears to describe the geometry of a four-dimensional superspace in
presence of a constant abelian supersymmetric gauge field background \cite%
{BGKL}. Subsequently in \cite{AI, CR2}, the Maxwell superalgebra has been
used to construct a pure supergravity action in four dimensions using
geometric methods. Further applications of the Maxwell superalgebra in the
context of supergravity can be found in \cite{PR, Ravera}. Such superalgebra
has the particularity of having more than one spinor charge and has been
obtained in \cite{AILW, CR1} through algebraic expansion mechanisms \cite%
{Sexp, AIPV}. More recently, a CS gravity action for a generalized Maxwell
superalgebra has been presented in \cite{CFRS, CFR} in three spacetime
dimensions. Although the CS supergravity theory is appropriately
constructed, it requires to consider a large amount of gauge fields whose
origin is due to the methodology. To our knowledge, the construction of a CS
supergravity action for the minimal Maxwell superalgebra in absence of extra
fields has not been presented yet. A well-defined minimal Maxwell
supergravity theory would allow to study more features of the
three-dimensional Maxwell supergravity, like its asymptotic structure and
non-relativistic limit.

In this paper, we present the minimal three-dimensional Maxwell CS
supergravity theory without considering a large amount of extra fields. In
particular, we present two alternative procedures to obtain it. We first
derive the minimal Maxwell superalgebra and the invariant tensor applying
the expansion method to a supersymmetric extension of the Lorentz algebra.
Then, we present the corresponding CS action, supersymmetry transformations
and field equations. In the second part, we recover the minimal Maxwell CS
supergravity as a flat limit of a minimal AdS-Lorentz CS supergravity.

The present work is organized as follows: In Section 2, we give a brief
review of Maxwell CS gravity theory in three spacetime dimensions. In
Section 3, we present a minimal Maxwell CS supergravity theory in three
dimensions. The Section 4 is devoted to the obtention of the Maxwell
supergravity theory through a flat limit of a minimal AdS-Lorentz CS
supergravity theory. In Section 5, we end our work with some comments about
future possible developments.

\section{Maxwell Chern-Simons gravity theory in 2+1 dimensions}

In this section, we briefly review the three-dimensional CS gravity theory
invariant under the Maxwell algebra \cite{SSV, HR, CCFRS, AFGHZ, CMMRSV}.
The explicit commutators of the Maxwell algebra can be obtained as a
deformation and enlargement of the Poincaré one. In particular, the Maxwell
algebra is spanned by the set $\left\{ J_{a},P_{a},Z_{a}\right\} $ whose
generators satisfy the following non-vanishing commutation relations:%
\begin{eqnarray}
\left[ J_{a},J_{b}\right] &=&\epsilon _{abc}J^{c}\,,\qquad \left[ J_{a},P_{b}%
\right] =\epsilon _{abc}P^{c}\,,  \notag \\
\left[ J_{a},Z_{b}\right] &=&\epsilon _{abc}Z^{c},\qquad\left[ P_{a},P_{b}%
\right] =\epsilon _{abc}Z^{c}\,,
\end{eqnarray}%
where $a,b,\dots =0,1,2$ are raised and lowered with the Minkowski metric $%
\eta _{ab}$ and $\epsilon _{abc}$ is the Levi-Civita tensor. Note that,
unlike the Poincaré algebra, the commutator of the translational generators $%
P_{a}$ is no longer zero but proportional to the new abelian generator $%
Z_{a} $.

In order to construct a CS action%
\begin{equation}
I_{CS}=\frac{k}{4\pi }\int_{\mathcal{M}}\left\langle AdA+\frac{2}{3}%
A^{3}\right\rangle \,,  \label{CSaction}
\end{equation}%
invariant under the Maxwell group, we require the Maxwell connection
one-form $A=A_{\mu }dx^{\mu }$ and the corresponding non-vanishing
components of the invariant tensor.

The gauge connection one-form for the Maxwell algebra reads%
\begin{equation}
A=\omega ^{a}J_{a}+e^{a}P_{a}+\sigma ^{a}Z_{a}\,,  \label{1f}
\end{equation}%
where $\omega ^{a}$ corresponds to the spin connection, $e^{a}$ is the
vielbein and $\sigma ^{a}$ is the gravitational Maxwell gauge field \cite%
{CMMRSV}. On the other hand, the non-vanishing components of the invariant
tensor for the Maxwell algebra are given by%
\begin{eqnarray}
\left\langle J_{a}J_{b}\right\rangle &=&\alpha _{0}\eta _{ab}\,,  \notag \\
\left\langle J_{a}P_{b}\right\rangle &=&\alpha _{1}\eta _{ab}\,,  \notag \\
\left\langle J_{a}Z_{b}\right\rangle &=&\alpha _{2}\eta _{ab}\,,  \label{it}
\\
\left\langle P_{a}P_{b}\right\rangle &=&\alpha _{2}\eta _{ab}\,,  \notag
\end{eqnarray}%
where $\alpha _{0}\,$, $\alpha _{1}$ and $\alpha _{2}$ are real constants.
In particular, the Maxwell algebra has the following quadratic Casimir \cite%
{Sorokas2, AFGHZ},%
\begin{equation}
C=\alpha _{0}J^{a}J_{a}+\alpha _{1}P^{a}J_{a}+\alpha _{2}\left(
P^{a}P_{a}+J^{a}Z_{a}\right) \,.
\end{equation}

Then considering the connection one-form (\ref{1f}) and the invariant tensor
(\ref{it}), the CS gravity action invariant under the Maxwell symmetry reads%
\begin{equation}
I_{CS}=\frac{k}{4\pi }\,\int_{M}\left[ \alpha _{0}\left( \omega ^{a}d\omega
_{a}+\frac{1}{3}\epsilon _{abc}\omega ^{a}\omega ^{b}\omega ^{c}\right)
+2\alpha _{1}R_{a}e^{a}+\alpha _{2}\left( T^{a}e_{a}+2R^{a}\sigma
_{a}\right) \right] \,,  \label{MCS}
\end{equation}%
where the Lorentz curvature and torsion two-forms are given respectively by%
\begin{eqnarray}
R^{a} &=&d\omega ^{a}+\frac{1}{2}\epsilon ^{abc}\omega _{b}\omega _{c}\,,
\notag \\
T^{a} &=&de^{a}+\epsilon ^{abc}\omega _{b}e_{c}\,.
\end{eqnarray}%
Here $k=\frac{1}{4G}$ is the CS level of the theory and it is related to the
gravitational constant $G$. Note that the term proportional to $\alpha _{0}$
is the exotic Lagrangian also known as the Lorentz Lagrangian. The second
term corresponds to the Einstein-Hilbert (EH) term while the term
proportional to $\alpha _{2}$ contains the explicit gravitational Maxwell
field $\sigma _{a}$. Note that each term is invariant under the Maxwell
symmetry. In particular, the local gauge transformations $\delta A=d\Lambda +%
\left[ A,\Lambda \right] $, with gauge parameter $\Lambda =\rho
^{a}J_{a}+\varepsilon ^{a}P_{a}+\gamma ^{a}Z_{a}$, are given by%
\begin{eqnarray}
\delta \omega ^{a} &=&D\rho ^{a}\,,  \notag \\
\delta e^{a} &=&D\varepsilon ^{a}-\epsilon ^{abc}\rho _{b}e_{c}\,, \\
\delta \sigma ^{a} &=&D\gamma ^{a}+\epsilon ^{abc}\left( e_{b}\varepsilon
_{b}-\rho _{b}\sigma _{c}\right) \,,  \notag
\end{eqnarray}%
where $Du^{a}=du^{a}+\epsilon ^{abc}\omega _{b}u_{c}$ is the Lorentz
covariant derivative.

The equations of motion derived from the action (\ref{MCS}) read%
\begin{eqnarray}
\delta \omega ^{a} &:&\qquad \alpha _{0}R_{a}+\alpha _{1}T_{a}+\alpha
_{2}\left( D\sigma _{a}+\frac{1}{2}\epsilon _{abc}e^{b}e^{c}\right) =0\,,
\notag \\
\delta e^{a} &:&\qquad \alpha _{1}R_{a}+\alpha _{2}T_{a}=0\,, \\
\delta \sigma ^{a} &:&\qquad \alpha _{2}R_{a}=0\,.  \notag
\end{eqnarray}%
In particular, when $\alpha _{2}\neq 0$, the field equations are given by
the vanishing of every curvature two-form%
\begin{eqnarray}
R^{a} &=&0\,,  \notag \\
T^{a} &=&0\,, \\
F^{a} &=&D\sigma _{a}+\frac{1}{2}\epsilon _{abc}e^{b}e^{c}=0\,.  \notag
\end{eqnarray}%
On the other hand, let us note that the EH dynamics is recovered in the
limit $\alpha _{2}=0$. Such interesting feature is proper of the Maxwell
like symmetry $\mathfrak{B}_{k}$ \cite{PSSS, CDMR} which corresponds to the
Maxwell algebra for $k=4$. As was shown in \cite{GRCS, CPRS1, CPRS2, CPRS3},
GR can be recovered under a particular limit from a CS and BI gravity theory
invariant under the Maxwell like groups.

\section{Minimal Maxwell Chern-Simons supergravity in three-dimensional
spacetime}

In this section, following a similar procedure used in \cite{CFRS}, we
present a novel way of finding the so called minimal Maxwell superalgebra $s%
\mathcal{M}$. Moreover the construction of the CS action invariant under
this superalgebra is presented. The minimal Maxwell superalgebra in $D=4$
dimensions was first introduced in \cite{BGKL}. \ Then, in $D=3$ dimensions,
a generalized minimal Maxwell superalgebra was derived as a semigroup
expansion ($S$-expansion) of the $\mathfrak{osp}(2|1)\otimes \mathfrak{sp}%
(2),$ and it was used in order to study the corresponding CS theory \cite%
{CFRS}. Although the procedure was useful to derive a three-dimensional CS
formalism, the generalized minimal Maxwell superalgebra obtained there
contains a large set of bosonic generators $\left\{ J_{ab},P_{a},Z_{ab},%
\tilde{Z}_{ab},\tilde{Z}_{a}\right\} $. In particular the bosonic content $%
\left\{ \tilde{Z}_{ab},\tilde{Z}_{a}\right\} $ implies a large number of
extra terms in the CS action whose presence is due to the considered
starting superalgebra.

Here we show that using a smaller structure as the original algebra, the
minimal Maxwell superalgebra can be obtained through the $S$-expansion \cite%
{Sexp}. As we shall see, the extra bosonic generators do not appear if a
super-Lorentz algebra is considered as the starting superalgebra. In
addition, such method allows not only to recover the complete set of
(anti-)commutation relations of the minimal Maxwell superalgebra but also
the invariant tensor which is required for the construction of a CS action.

The most natural minimal supersymmetric extension of the Lorentz algebra in
three spacetime dimensions is spanned by Lorentz generators $M_{a}$ and
Majorana fermionic generators $Q_{\alpha } \ (\alpha =1,2)$ \cite{LuNo}. The
(anti-)commutation relations read%
\begin{eqnarray}
\left[ M_{a},M_{b}\right] &=&\epsilon _{abc}M^{c}\,,  \notag \\
\left[ M_{a},Q_{\alpha }\right] &=&\frac{1}{2}\,\left( \Gamma _{a}\right) _{%
\text{ }\alpha }^{\beta }Q_{\beta }\,,  \label{Lorentz} \\
\left\{ Q_{\alpha },Q_{\beta }\right\} &=&-\frac{1}{2}\,\left( C\Gamma
^{a}\right) _{\alpha \beta }M_{a}\,,  \notag
\end{eqnarray}%
where $a,b,\dots =0,1,2$ are the Lorentz indices, $\Gamma _{a}$ are the
Dirac matrices in three dimensions, and $C$ represents the charge
conjugation matrix,%
\begin{equation}
C_{\alpha \beta }=C^{\alpha \beta }=%
\begin{pmatrix}
0 & -1 \\
1 & 0%
\end{pmatrix}%
\,.
\end{equation}%
One can easily check the consistency of the (anti-)commutation relations of
the super-Lorentz algebra $s\mathcal{L}$ (\ref{Lorentz}) by checking the
Jacobi identities, and by considering that $C^{T}=-C$ and $C\Gamma
^{a}=(C\Gamma ^{a})^{T}$.

Let $S_{E}^{\left( 4\right) }=\left\{ \lambda _{0},\lambda _{1},\lambda
_{2},\lambda _{3},\lambda _{4},\lambda _{5}\right\} $ be the abelian
semigroup whose elements satisfy the multiplication law%
\begin{equation}
\lambda _{\alpha }\lambda _{\beta }=\left\{
\begin{array}{c}
\lambda _{\alpha +\beta }\,\text{, \ \ \ \ when }\alpha +\beta \leq 5\,, \\
\lambda _{5}\,\text{, \ \ \ \ \ \ \ \ when }\alpha +\beta >5\,,%
\end{array}%
\right.
\end{equation}%
where $\lambda _{5}=0_{s}$ is the zero element of the semigroup.

Following the definitions of \cite{Sexp}, after extracting a resonant
subalgebra of $S_{E}^{\left( 4\right) }\times s\mathcal{L}$ and applying its
$0_{s}$-reduction, one finds a new superalgebra whose generators $\left\{
J_{a},P_{a},Z_{a},Q_{\alpha },\Sigma _{\alpha }\right\} $ are related to the
super-Lorentz ones as%
\begin{equation}
\begin{tabular}{lll}
$J_{a}=\lambda _{0}M_{a}\,,$ & $\ell P_{a}=\lambda _{2}M_{a}\,,$ & $\ell
^{2}Z_{a}=\lambda _{4}M_{a}\,,$ \\
$\ell ^{1/2}Q_{\alpha }=\lambda _{1}Q_{\alpha }\,,\,$ & $\ell ^{3/2}\Sigma
_{\alpha }=\lambda _{3}Q_{\alpha }\,.$ &
\end{tabular}%
\end{equation}%
Using the multiplication law of the semigroup and the super-Lorentz
(anti-)commutators, one can see that the expanded generators satisfy the
following non-vanishing (anti-)commutation relations%
\begin{eqnarray}
\left[ J_{a},J_{b}\right] &=&\epsilon _{abc}J^{c}\,,\qquad \left[ J_{a},P_{b}%
\right] =\epsilon _{abc}P^{c}\,,  \notag \\
\left[ J_{a},Z_{b}\right] &=&\epsilon _{abc}Z^{c}\,,\qquad\left[ P_{a},P_{b}%
\right] =\epsilon _{abc}Z^{c}\,,  \notag \\
\left[ J_{a},Q_{\alpha }\right] &=&\frac{1}{2}\,\left( \Gamma _{a}\right) _{%
\text{ }\alpha }^{\beta }Q_{\beta }\,,\text{ \ \ }  \notag \\
\left[ J_{a},\Sigma _{\alpha }\right] &=&\frac{1}{2}\,\left( \Gamma
_{a}\right) _{\text{ }\alpha }^{\beta }\Sigma _{\beta }\,,\text{ \ \ }
\label{sp1} \\
\left[ P_{a},Q_{\alpha }\right] &=&\frac{1}{2}\,\left( \Gamma _{a}\right) _{%
\text{ }\alpha }^{\beta }\Sigma _{\beta }\,,\text{ }  \notag \\
\left\{ Q_{\alpha },Q_{\beta }\right\} &=&-\frac{1}{2}\left( C\Gamma
^{a}\right) _{\alpha \beta }P_{a}\,,  \notag \\
\left\{ Q_{\alpha },\Sigma _{\beta }\right\} &=&-\frac{1}{2}\,\left( C\Gamma
^{a}\right) _{\alpha \beta }Z_{a}\,.  \notag
\end{eqnarray}%
The (anti-)commutation relations (\ref{sp1}) correspond to the minimal
Maxwell superalgebra $s\mathcal{M}$, introduced in \cite{BGKL2} in three
spacetime dimensions. This supersymmetric extension of the Maxwell symmetry
is characterized by the introduction of a new Majorana spinor charge $\Sigma
$ which is required to satisfy the Jacobi identity $\left( P,Q,Q\right) $.
The presence of a second abelian spinorial generators is not new in the
literature and has already been studied in the context of $D=11$
supergravity \cite{AF} and superstring theory \cite{Green}.

The non-vanishing components of the invariant tensor of the Maxwell
superalgebra can be obtained in terms of the super-Lorentz ones using the
definitions of the $S$-expansion method,%
\begin{eqnarray}
\left\langle J_{a}J_{b}\right\rangle &=&\mu _{0}\left\langle
M_{a}M_{b}\right\rangle _{s\mathcal{L}}=\mu _{0}\eta _{ab}\,,  \notag \\
\left\langle J_{a}P_{b}\right\rangle &=&\frac{\mu _{2}}{\ell }\left\langle
M_{a}M_{b}\right\rangle _{s\mathcal{L}}=\frac{\mu _{2}}{\ell }\eta _{ab}\,,
\notag \\
\left\langle J_{a}Z_{b}\right\rangle &=&\frac{\mu _{4}}{\ell ^{2}}%
\left\langle M_{a}M_{b}\right\rangle _{s\mathcal{L}}=\frac{\mu _{4}}{\ell
^{2}}\eta _{ab}\,,  \notag \\
\left\langle P_{a}P_{b}\right\rangle &=&\frac{\mu _{4}}{\ell ^{2}}%
\left\langle M_{a}M_{b}\right\rangle _{s\mathcal{L}}=\frac{\mu _{4}}{\ell
^{2}}\eta _{ab}\,, \\
\left\langle Q_{\alpha }Q_{\beta }\right\rangle &=&\frac{\mu _{2}}{\ell }%
\left\langle Q_{\alpha }Q_{\beta }\right\rangle _{s\mathcal{L}}=\frac{\mu
_{2}}{\ell }C_{\alpha \beta }\,,  \notag \\
\left\langle Q_{\alpha }\Sigma _{\beta }\right\rangle &=&\frac{\mu _{4}}{%
\ell ^{2}}\left\langle Q_{\alpha }Q_{\beta }\right\rangle _{s\mathcal{L}}=%
\frac{\mu _{4}}{\ell ^{2}}C_{\alpha \beta }\,,  \notag
\end{eqnarray}%
where $\mu _{0},\mu _{2}$ and $\mu _{4}$ are arbitrary constants. For
convenience, we will redefine the $\mu $'s as follows%
\begin{equation*}
\mu _{0}\rightarrow \alpha _{0},\text{ \ \ }\mu _{2}\rightarrow \ell \alpha
_{1},\text{ \ \ }\mu _{4}\rightarrow \ell ^{2}\alpha _{2}\,,
\end{equation*}%
and the invariant tensor takes de the form%
\begin{eqnarray}
\left\langle J_{a}J_{b}\right\rangle &=&\alpha _{0}\eta _{ab}\,,\text{\qquad
}\left\langle P_{a}P_{b}\right\rangle =\alpha _{2}\eta _{ab}\,,  \notag \\
\left\langle J_{a}P_{b}\right\rangle &=&\alpha _{1}\eta _{ab}\,,\text{\qquad
}\left\langle Q_{\alpha }Q_{\beta }\right\rangle =\alpha _{1}C_{\alpha \beta
}\,,  \label{InvTensMax} \\
\left\langle J_{a}Z_{b}\right\rangle &=&\alpha _{2}\eta _{ab}\,,\text{\qquad
}\left\langle Q_{\alpha }\Sigma _{\beta }\right\rangle =\alpha _{2}C_{\alpha
\beta }\,.  \notag
\end{eqnarray}

The connection one-form reads
\begin{equation}
A=\omega ^{a}J_{a}+e^{a}P_{a}+\sigma ^{a}Z_{a}+\bar{\psi}Q+\bar{\xi}\Sigma
\,,  \label{1fP}
\end{equation}%
where $\omega ^{a}$ is the spin connection one-form, $e^{a}$ corresponds to
the vielbein one-form, $\sigma ^{a}$ is the Maxwell gravity field one-form
while $\psi $ and\ $\xi $ are fermionic fields.

The corresponding curvature two-form is given by%
\begin{equation}
F=R^{a}J_{a}+\mathcal{T}^{a}P_{a}+\mathcal{F}^{a}Z_{a}+\nabla \bar{\psi}%
Q+\nabla \bar{\xi}\Sigma \,,
\end{equation}%
with%
\begin{eqnarray}
R^{a} &=&d\omega ^{a}+\frac{1}{2}\epsilon ^{abc}\omega _{b}\omega _{c}\,,
\notag \\
\mathcal{T}^{a} &=&de^{a}+\epsilon ^{abc}\omega _{b}e_{c}+\frac{1}{4}\bar{%
\psi}\Gamma ^{a}\psi \,,  \label{bosc} \\
\mathcal{F}^{a} &=&d\sigma ^{a}+\epsilon ^{abc}\omega _{b}\sigma _{c}+\frac{1%
}{2}\epsilon _{\text{ }}^{abc}e_{b}e_{c}+\frac{1}{2}\bar{\psi}\Gamma ^{a}\xi
\,.  \notag
\end{eqnarray}%
Here, the covariant derivative $\nabla =d+[A,\cdot ]$ acting on spinors read%
\begin{eqnarray}
\nabla \psi &=&d\psi +\frac{1}{2}\,\omega ^{a}\Gamma _{a}\psi \,,  \notag \\
\nabla \xi &=&d\xi +\frac{1}{2}\,\omega ^{a}\Gamma _{a}\xi \,+\frac{1}{2}%
\,e^{a}\Gamma _{a}\psi \,.  \label{ferc}
\end{eqnarray}%
The CS supergravity action can be written considering the non-vanishing
component of the invariant tensor (\ref{InvTensMax}) and the gauge
connection one-form (\ref{1fP}),%
\begin{eqnarray}
I_{s\mathcal{M}} &=&\frac{k}{4\pi }\int \alpha _{0}\left( \,\omega
^{a}d\omega _{a}+\frac{1}{3}\,\epsilon _{abc}\omega ^{a}\omega ^{b}\omega
^{c}\right)  \notag \\
&&+\alpha _{1}\left( 2e^{a}R_{a}-\bar{\psi}\nabla \psi \right) \,+\alpha
_{2}\left( 2R^{a}\sigma _{a}+e^{a}T_{a}-\bar{\psi}\nabla \xi -\bar{\xi}%
\nabla \psi \right) \,,  \label{sMCS}
\end{eqnarray}%
where $T^{a}=De^{a}$ is the usual torsion two-form.$\,$

Note that the term proportional to $\alpha _{0}$ contains just the exotic
Lagrangian. The piece along $\alpha _{1}$ contains the EH and the
Rarita-Schwinger terms while the term proportional to $\alpha _{2}$ is of
particular interest since it contains the Maxwell gravitational field $%
\sigma ^{a}$ plus a torsional term. Let us note that the fermionic terms
contribute only to the $\alpha _{1}$ and $\alpha _{2}$ sectors of the
bosonic action (\ref{MCS}). In addition, the CS action (\ref{sMCS})
reproduces the pure three-dimensional supergravity action when the exotic CS
terms are neglected ($\alpha _{0}=\alpha _{2}=0$). Such feature also appears
on four-dimensional supergravity action based on the Maxwell supergroup \cite%
{CR2} using the MacDowell-Mansouri formalism \cite{MM}.

By construction, the CS action (\ref{sMCS}) is invariant under the gauge
transformation $\delta A=d\Lambda +\left[ A,\Lambda \right] $. In
particular, the action is invariant under the following local supersymmetry
transformation laws%
\begin{eqnarray}
\delta \omega ^{a} &=&0\,,  \notag \\
\delta e^{a} &=&\frac{1}{2}\,\bar{\epsilon}\Gamma ^{a}\psi \,,  \notag \\
\delta \sigma ^{a} &=&\frac{1}{2}\,\bar{\epsilon}\Gamma ^{a}\xi +\frac{1}{2}%
\,\bar{\varrho}\Gamma ^{a}\psi \,,  \label{Msusy} \\
\delta \psi &=&d\epsilon +\frac{1}{2}\,\omega ^{a}\Gamma _{a}\epsilon \,,
\notag \\
\delta \xi &=&d\varrho +\frac{1}{2}\,\omega ^{a}\Gamma _{a}\varrho \,+\frac{1%
}{2}e^{a}\Gamma _{a}\epsilon \,,  \notag
\end{eqnarray}%
where the gauge parameter is $\Lambda =\rho ^{a}J_{a}+\varepsilon
^{a}P_{a}+\gamma ^{a}Z_{a}+\bar{\epsilon}Q+\bar{\varrho}\Sigma $.

The equations of motion derived from (\ref{MCS}) are%
\begin{eqnarray}
\delta \omega ^{a} &:&\qquad \alpha _{0}R_{a}+\alpha _{1}\mathcal{T}%
_{a}+\alpha _{2}\mathcal{F}_{a}=0\,,  \notag \\
\delta e^{a} &:&\qquad \alpha _{1}R_{a}+\alpha _{2}\mathcal{T}_{a}=0\,,
\notag \\
\delta \sigma ^{a} &:&\qquad \alpha _{2}R_{a}=0\,,  \label{meq} \\
\delta \bar{\psi} &:&\qquad \alpha _{1}\nabla \psi +\alpha _{2}\nabla \xi
=0\,,  \notag \\
\delta \bar{\xi} &:&\qquad \alpha _{2}\nabla \psi =0\,.  \notag
\end{eqnarray}%
As in the bosonic case, the field equations reduce to the vanishing of the
curvature two-forms provided $\alpha _{2}\neq 0$,%
\begin{equation}
\begin{tabular}{lll}
$R^{a}=0\,,$ & $\mathcal{T}^{a}=0\,,$ & $\mathcal{F}^{a}=0\,,$ \\
$\nabla \psi =0\,,$ & $\nabla \xi =0\,.$ &
\end{tabular}%
\end{equation}%
As was recently shown in the Maxwell gravity case \cite{CMMRSV}, $\sigma
^{a} $ is not simply an extra field but modifies the asymptotic charges of
the solutions. It would be interesting to extend the results obtained in
\cite{CMMRSV}\ to the Maxwell supergravity theory presented here. As in the
bosonic theory, one could expect a deformation of the super $\mathfrak{bms}%
_{3}$ algebra \cite{BM, BDMT} as the corresponding asymptotic symmetry. In
particular, as in the finite Maxwell superalgebra, two infinite-dimensional
fermionic generators should appear in the asymptotic structure.

As an ending remark, the CS theory presented here offers us an alternative
three-dimensional minimal supergravity with vanishing cosmological constant.
Then, analogously to the Poincaré CS supergravity which appears as a flat
limit of the AdS one, it is natural to expect that the Maxwell CS
supergravity obtained here can also be found as a flat limit of a particular
supergravity theory. The next section is devoted to the obtention of the
Maxwell supergravity theory as a flat limit of an AdS-Lorent supergravity.

\section{Maxwell supergravity theory as a flat limit of the AdS-Lorentz
supergravity}

It has been shown that the Maxwell symmetry can be alternatively obtained as
an IW contraction of an enlarged symmetry denoted as the AdS-Lorentz algebra
\cite{PSSS}. Such algebra and its generalizations have been useful to relate
diverse (pure) Lovelock gravity theories \cite{CDIMR, CMR, CR3}. At the
supersymmetric level, numerous studies have been done in four and three
dimensions leading to interesting AdS-Lorentz supergravities \cite{FISV,
CRS, CIRR, BR}. Recently, it was shown in three dimensions that a
generalized Maxwell supergravity can be obtained as an IW contraction of a
generalized AdS-Lorentz supergravity model \cite{CFR}. Nevertheless, as was
discussed in the previous section, the field content of such theories is
large and the physical interpretation of the extra fields remains unknown.
In particular, the extra fields related to the set of generators $\left\{
\tilde{Z}_{ab},\tilde{Z}_{a}\right\} $ appear as a consequence of the
procedure used to construct the supergravity theories.

In this section we show that the minimal Maxwell supergravity theory
presented previously can be recovered as a flat limit of a minimal
AdS-Lorentz supergravity. To this purpose, let us first consider a novel
supersymmetric extension of the AdS-Lorentz algebra in three dimensions. We
extend the AdS-Lorentz algebra generated by $\{J_{a},P_{a},Z_{a}\}$ with the
fermionic generators $\{Q_{\alpha },\Sigma _{\alpha }\}$, and we get the
following superalgebra%
\begin{eqnarray}
\left[ J_{a},J_{b}\right] &=&\epsilon _{abc}J^{c}\,,\qquad \ \ \ \left[
J_{a},P_{b}\right] =\epsilon _{abc}P^{c}\,,  \notag \\
\left[ P_{a},P_{b}\right] &=&\epsilon _{abc}Z^{c}\,,\qquad \ \ \ \left[
J_{a},Z_{b}\right] =\epsilon _{abc}Z^{c}\,, \label{bosAdS-L}\\
\left[ Z_{a},Z_{b}\right] &=&\frac{1}{\ell ^{2}}\epsilon
_{abc}Z^{c}\,,\qquad \left[ P_{a},Z_{b}\right] =\frac{1}{\ell ^{2}}\epsilon
_{abc}P^{c}\,,  \notag
\end{eqnarray}%
\begin{eqnarray}
\left[ J_{a},Q_{\alpha }\right] &=&\frac{1}{2}\,\left( \Gamma _{a}\right) _{%
\text{ }\alpha }^{\beta }Q_{\beta }\,,\qquad \ \,\left[ J_{a},\Sigma
_{\alpha }\right] =\frac{1}{2}\,\left( \Gamma _{a}\right) _{\text{ }\alpha
}^{\beta }\Sigma _{\beta }\,,  \notag \\
\left[ P_{a},Q_{\alpha }\right] &=&\frac{1}{2}\,\,\left( \Gamma _{a}\right)
_{\text{ }\alpha }^{\beta }\Sigma _{\beta }\,,\qquad \ \,\left[ P_{a},\Sigma
_{\alpha }\right] =\frac{1}{2\ell ^{2}}\,\left( \Gamma _{a}\right) _{\text{ }%
\alpha }^{\beta }Q_{\beta }\,,  \notag \\
\left[ Z_{a},Q_{\alpha }\right] &=&\frac{1}{2\ell ^{2}}\,\left( \Gamma
_{a}\right) _{\text{ }\alpha }^{\beta }Q_{\beta }\,,\qquad \left[
Z_{a},\Sigma _{\alpha }\right] =\frac{1}{2\ell ^{2}}\,\left( \Gamma
_{a}\right) _{\text{ }\alpha }^{\beta }\Sigma _{\beta }\,, \label{ferAdS-L}\\
\left\{ Q_{\alpha },Q_{\beta }\right\} &=&-\frac{1}{2}\,\left( C\Gamma
^{a}\right) _{\alpha \beta }P_{a}\,,\text{ \ \ }\left\{ Q_{\alpha },\Sigma
_{\beta }\right\} =-\frac{1}{2}\,\left( C\Gamma ^{a}\right) _{\alpha \beta
}Z_{a}\,,\,  \notag \\
\left\{ \Sigma _{\alpha },\Sigma _{\beta }\right\} &=&-\frac{1}{2\ell ^{2}}%
\,\,\left( C\Gamma ^{a}\right) _{\alpha \beta }P_{a}\,.  \notag
\end{eqnarray}%
We will refer to this superalgebra as the minimal AdS-Lorentz superalgebra
in three dimensions. Note that the supersymmetric extension of the
AdS-Lorentz algebra is not unique. However, to our knowledge, this
corresponds to the minimal supersymmetric extension of the AdS-Lorentz
algebra containing two spinor charges.

One can see that the limit $\ell \rightarrow \infty $ leads appropriately to
the minimal Maxwell superalgebra (\ref{sp1}) considered in the previous
section. As we shall see, the flat limit can also be directly performed at
the level of the action, supersymmetry transformations and field equations.

In order to write down a CS action for the minimal AdS-Lorentz superalgebra,
let us consider the connection one-form
\begin{equation}
A=\omega ^{a}J_{a}+e^{a}P_{a}+\sigma ^{a}Z_{a}+\bar{\psi}Q+\bar{\xi}\Sigma
\,,  \label{adslor con}
\end{equation}%
and the corresponding curvature two-form%
\begin{equation}
F=R^{a}J_{a}+\mathcal{T}^{a}P_{a}+\mathcal{F}^{a}Z_{a}+\nabla \bar{\psi}%
Q+\nabla \bar{\xi}\Sigma \,,
\end{equation}%
where%
\begin{eqnarray}
R^{a} &=&d\omega ^{a}+\frac{1}{2}\epsilon ^{abc}\omega _{b}\omega _{c}\,,
\notag \\
\mathcal{T}^{a} &=&de^{a}+\epsilon ^{abc}\omega _{b}e_{c}+\frac{1}{\ell ^{2}}%
\epsilon ^{abc}\sigma _{b}e_{c}+\frac{1}{4}\bar{\psi}\Gamma ^{a}\psi +\frac{1%
}{4\ell ^{2}}\bar{\xi}\Gamma ^{a}\xi \,, \\
\mathcal{F}^{a} &=&d\sigma ^{a}+\epsilon ^{abc}\omega _{b}\sigma _{c}+\frac{1%
}{2\ell ^{2}}\epsilon ^{abc}\sigma _{b}\sigma _{c}+\frac{1}{2}\epsilon
^{abc}e_{b}e_{c}+\frac{1}{2}\bar{\psi}\Gamma ^{a}\xi \,.  \notag
\end{eqnarray}%
Furthermore,
\begin{eqnarray}
\nabla \psi &=&d\psi +\frac{1}{2}\,\omega ^{a}\Gamma _{a}\psi +\frac{1}{%
2\ell ^{2}}\,\sigma ^{a}\Gamma _{a}\psi +\frac{1}{2\ell ^{2}}\,e^{a}\Gamma
_{a}\xi \,,  \notag \\
\nabla \xi &=&d\xi +\frac{1}{2}\,\omega ^{a}\Gamma _{a}\xi \,+\frac{1}{2\ell
^{2}}\,\sigma ^{a}\Gamma _{a}\xi +\frac{1}{2}\,e^{a}\Gamma _{a}\psi \,.
\end{eqnarray}%
Note that the curvatures reduce to the Maxwell ones (\ref{bosc}) and (\ref%
{ferc}) by performing the limit $\ell \rightarrow \infty $.

The non-vanishing components of the invariant tensor are given by (\ref%
{InvTensMax}) along with
\begin{align}
\left\langle P_{a}Z_{b}\right\rangle & =\frac{\alpha _{1}}{\ell ^{2}}\eta
_{ab}\,,  \notag \\
\left\langle Z_{a}Z_{b}\right\rangle & =\frac{\alpha _{2}}{\ell ^{2}}\eta
_{ab}\,,  \label{AdSLorInv} \\
\left\langle \Sigma _{\alpha }\Sigma _{\beta }\right\rangle & =\frac{\alpha
_{1}}{\ell ^{2}}C_{\alpha \beta }\,.  \notag
\end{align}
Thus, the most general quadratic Casimir of the minimal AdS-Lorentz
superalgebra is
\begin{equation}
C=\alpha_{0}J^{a}J_{a}+\alpha_{1}\left(P^{a}J_{a}+\frac{1}{\ell^{2}}%
P^{a}Z_{a}+\bar{Q}Q+\frac{1}{\ell^{2}}\bar{\Sigma}\Sigma\right)+\alpha_2%
\left(P^{a}P_{a}+J^{a}Z_{a}+\frac{1}{\ell^{2}}Z^{a}Z_{a}+\bar{Q}%
\Sigma\right)\,.
\end{equation}
Let us note that the flat limit $\ell \rightarrow \infty$ at the level of
the invariant tensor of the AdS-Lorentz superalgebra leads us to the
non-degenerate invariant tensor of the minimal Maxwell superalgebra (\ref%
{InvTensMax}). Then, using the connection one-form (\ref{adslor con}) and
the invariant tensors (\ref{InvTensMax}) and (\ref{AdSLorInv}) in the CS
action (\ref{CSaction}), we get
\begin{align}
I_{sAdS-\mathcal{L}}\,=\,& \frac{k}{4\pi }\int \alpha _{0}\left( \,\omega
^{a}d\omega _{a}+\frac{1}{3}\,\epsilon _{abc}\omega ^{a}\omega ^{b}\omega
^{c}\right)  \notag \\
& +\alpha _{1}\left( 2e^{a}R_{a}+\frac{2}{\ell ^{2}}\,e^{a}F_{a}+\frac{1}{%
3\ell ^{2}}\,\epsilon _{abc}e^{a}e^{b}e^{c}-\bar{\psi}\nabla \psi -\frac{1}{%
\ell ^{2}}\,\bar{\xi}\nabla \xi \right)  \notag \\
& +\alpha _{2}\left( 2R^{a}\sigma _{a}+\frac{2}{\ell ^{2}}\,F^{a}\sigma
_{a}+e^{a}T_{a}+\frac{1}{\ell ^{2}}\,\epsilon _{abc}e^{a}\sigma ^{b}e^{c}-%
\bar{\psi}\nabla \xi -\bar{\xi}\nabla \psi \right) \,,  \label{AdSLorAct}
\end{align}%
where $T^{a}=de^{a}+\epsilon ^{abc}\omega _{b}e_{c}$ is the torsion two-form
and%
\begin{equation}
F^{a}=d\sigma ^{a}+\epsilon ^{abc}\omega _{b}\sigma _{c}+\frac{1}{2\ell ^{2}}%
\epsilon ^{abc}\sigma _{b}\sigma _{c}\,.
\end{equation}%
Although the field content of the AdS-Lorentz supergravity theory is the
same as in the super Maxwell one, the CS action is much richer. As in the
four-dimensional case \cite{CRS}, this symmetry allows the presence of a
cosmological constant term which was absent in the super Maxwell theory.
This is due to the presence of the component $\left\langle
P_{a}Z_{b}\right\rangle $ of the invariant tensor which vanishes in the
Maxwell case. Interestingly, from the minimal\ AdS-Lorentz supergravity
action (\ref{AdSLorAct}) we see that the vanishing cosmological constant
limit $\ell \rightarrow \infty $ leads us to the minimal Maxwell
supergravity action (\ref{sMCS}) introduced in the previous section. This is
very similar to the flat behavior present in AdS supergravity and in the
bosonic AdS-Lorentz gravity theory. For completeness we apply the flat limit
at the level of the supersymmetry transformations and equations of motion.

The CS action (\ref{AdSLorAct}) is invariant by construction under the gauge
transformation $\delta A=d\Lambda +\left[ A,\Lambda \right] $. In
particular, the supersymmetry transformation laws read%
\begin{eqnarray}
\delta \omega ^{a} &=&0\,,  \notag \\
\delta e^{a} &=&\frac{1}{2}\,\bar{\epsilon}\Gamma ^{a}\psi \,+\frac{1}{2\ell
^{2}}\,\bar{\varrho}\Gamma ^{a}\xi ,  \notag \\
\delta \sigma ^{a} &=&\frac{1}{2}\,\bar{\epsilon}\Gamma ^{a}\xi +\frac{1}{2}%
\,\bar{\varrho}\Gamma ^{a}\psi \,,  \label{Asusy} \\
\delta \psi &=&d\epsilon +\frac{1}{2}\,\omega ^{a}\Gamma _{a}\epsilon +\frac{%
1}{2\ell ^{2}}\sigma ^{a}\Gamma _{a}\epsilon +\frac{1}{2\ell ^{2}}%
e^{a}\Gamma _{a}\varrho \,,  \notag \\
\delta \xi &=&d\varrho +\frac{1}{2}\,\omega ^{a}\Gamma _{a}\varrho \,+\frac{1%
}{2}e^{a}\Gamma _{a}\epsilon +\frac{1}{2\ell ^{2}}\sigma ^{a}\Gamma
_{a}\varrho \,,  \notag
\end{eqnarray}%
with gauge parameter $\Lambda =\rho ^{a}J_{a}+\varepsilon ^{a}P_{a}+\gamma
^{a}Z_{a}+\bar{\epsilon}Q+\bar{\varrho}\Sigma $. Note that the supersymmetry
transformations (\ref{Asusy}) reduce to the Maxwell ones of eq. (\ref{Msusy}%
) in the limit $\ell \rightarrow \infty $.

The field equations are given by
\begin{eqnarray}
\delta \omega ^{a} &:&\text{ \ \ \ }\alpha _{0}R_{a}+\alpha _{1}\mathcal{T}%
_{a}+\alpha _{2}\mathcal{F}_{a}\,=0\,,  \notag \\
\delta e^{a} &:&\text{ \ \ \ }\alpha _{1}\left( R_{a}+\frac{1}{\ell ^{2}}\,%
\mathcal{F}_{a}\right) +\alpha _{2}\mathcal{T}_{a}=0\,,  \notag \\
\delta \sigma ^{a} &:&\text{ \ \ }\frac{\alpha _{1}}{\ell ^{2}}\,\mathcal{T}%
_{a}+\alpha _{2}\left( R_{a}+\frac{1}{\ell ^{2}}\,\mathcal{F}_{a}\right)
\,=0\,,  \label{adsleq} \\
\delta \bar{\psi} &:&\qquad \alpha _{1}\nabla \psi +\alpha _{2}\nabla \xi
=0\,,  \notag \\
\delta \bar{\xi} &:&\qquad \alpha _{2}\nabla \psi +\frac{\alpha _{1}}{\ell
^{2}}\nabla \xi =0\,.  \notag
\end{eqnarray}%
When $\alpha _{0}$, $\alpha _{1}$ and $\alpha _{2}$ are independent, the
equations of motion reduce to the vanishing of the curvature two-forms,%
\begin{equation}
\begin{tabular}{lll}
$R^{a}=0\,,$ & $\mathcal{T}^{a}=0\,,$ & $\mathcal{F}^{a}=0\,,$ \\
$\nabla \psi =0\,,$ & $\nabla \xi =0\,.$ &
\end{tabular}%
\end{equation}%
It is simple to verify that the vanishing cosmological constant limit $\ell
\rightarrow \infty $ in (\ref{adsleq}) leads to the super Maxwell field
equations (\ref{meq}).

It is important to clarify that an alternative supersymmetric extension of
the AdS-Lorentz algebra can be obtained with one spinor generator $\tilde{Q}%
_{\alpha }$ such that%
\begin{eqnarray}
\left[ \tilde{J}_{a},\tilde{J}_{b}\right]  &=&\epsilon _{abc}\tilde{J}%
^{c}\,,\qquad \ \ \ \left[ \tilde{J}_{a},\tilde{P}_{b}\right] =\epsilon
_{abc}\tilde{P}^{c}\,,  \notag \\
\left[ \tilde{P}_{a},\tilde{P}_{b}\right]  &=&\epsilon _{abc}\tilde{Z}%
^{c}\,,\qquad \ \ \ \left[ \tilde{J}_{a},\tilde{Z}_{b}\right] =\epsilon
_{abc}\tilde{Z}^{c}\,,  \notag \\
\left[ \tilde{Z}_{a},\tilde{Z}_{b}\right]  &=&\frac{1}{\ell ^{2}}\epsilon
_{abc}\tilde{Z}^{c}\,,\qquad \left[ \tilde{P}_{a},\tilde{Z}_{b}\right] =%
\frac{1}{\ell ^{2}}\epsilon _{abc}\tilde{P}^{c}\,,  \notag \\
\left[ \tilde{J}_{a},\tilde{Q}_{\alpha }\right]  &=&\frac{1}{2}\,\left(
\Gamma _{a}\right) _{\text{ }\alpha }^{\beta }\tilde{Q}_{\beta }\,, \label{sAdS-L} \\
\left[ \tilde{P}_{a},\tilde{Q}_{\alpha }\right]  &=&\frac{1}{2}\,\,\left(
\Gamma _{a}\right) _{\text{ }\alpha }^{\beta }\tilde{Q}_{\beta }\,, \notag \\
\left[ \tilde{Z}_{a},\tilde{Q}_{\alpha }\right]  &=&\frac{1}{2\ell ^{2}}%
\,\left( \Gamma _{a}\right) _{\text{ }\alpha }^{\beta }\tilde{Q}_{\beta }\,,
\notag \\
\left\{ \tilde{Q}_{\alpha },\tilde{Q}_{\beta }\right\}  &=&-\frac{1}{2}%
\,\left( C\Gamma ^{a}\right) _{\alpha \beta }\tilde{Z}_{a}-\frac{1}{2}%
\,\left( C\Gamma ^{a}\right) _{\alpha \beta }\tilde{P}_{a}\,.  \notag
\end{eqnarray}%
This superalgebra can be seen as a supersymmetric extension of a semisimple
generalization of the Poincaré algebra \cite{SS}. Although this superalgebra
contains the bosonic AdS-Lorentz subalgebra, its supersimmetrization is
quite different from the minimal one presented previously (see eqs. (\ref{bosAdS-L}%
)-(\ref{ferAdS-L})). One can see that there is no redefinition of the generators allowing to
relate them. Moreover, a reinterpretation of the $P_{a}$ and $Z_{a}$ generators would
modify drastically the bosonic algebra. More recently, a three-dimensional CS supergravity action
invariant under this super AdS-Lorentz algebra has been presented in \cite%
{FISV} allowing to introduce a generalized cosmological term to the EH term.

In particular, a non-standard Maxwell superalgebra \cite{Sorokas2, Lukierski}
can be obtained applying the IW contraction to \ref{sAdS-L}. Although a supergravity CS
action can be constructed from this alternative super AdS-Lorentz version,
the IW contraction does not reproduces a supergravity theory. Indeed, in the
non-standard Maxwell superalgebra we have
\begin{equation}
\left\{ \tilde{Q}_{\alpha },\tilde{Q}_{\beta }\right\} =-\frac{1}{2}\,\left(
C\Gamma ^{a}\right) _{\alpha \beta }\tilde{Z}_{a}\,.
\end{equation}%
Here, the four-momentum generators $\tilde{P}_{a}$ are no more expressed as
bilinear expressions of the fermionic generators $\tilde{Q}$ which leads to
an exotic supersymmetric action.

\section{Discussion}

The minimal supersymmetric extension of the three-dimensional Maxwell
Chern-Simons gravity has been presented by expanding a Lorentz superalgebra.
The Maxwell superalgebra obtained is the minimal one presented in \cite{BGKL}
containing two spinor generators $Q_{\alpha }$ and $\Sigma _{\alpha }$. The
methodology considered here allows us to construct the CS supergravity
action invariant under the minimal superMaxwell which is characterized by
three coupling constants. In particular, the gravitational Maxwell field $%
\sigma _{a}$ appears in the $\alpha _{2}$ sector. In addition, the
supergravity theory presented here does not contain additional bosonic
fields as those appearing in the generalized Maxwell superalgebra \cite%
{CFRS, CFR}.

Interestingly, the present Maxwell supergravity theory has also been
obtained by applying a vanishing cosmological constant limit to a minimal
AdS-Lorentz supergravity theory. The flat limit was presented not only in
the commutation relations but also in the action, field equations and
supersymmetry transformation laws. This flat behavior appears also at the
bosonic level, including its infinite-dimensional symmetries \cite{CCRS,
CMMRSV} and higher-spin extension of the Maxwell and AdS-Lorentz gravity
\cite{CCFRS}.

Having a well defined Maxwell supergravity theory in three spacetime
dimensions, it would be interesting to go further and analyze the influence
of the gravitational Maxwell field in the asymptotic symmetry of this
supergravity theory. Analogously to the bosonic case, one could expect a
deformation and enlargement of the super $\mathfrak{bms}_{3}$ algebra [work
in progress]. One could go even further and study the asymptotic symmetry of
the AdS-Lorentz supergravity in three dimensions and analyze the existence
of a flat limit.

Another important aspect that deserves further investigation is the study of
the non-relativistic (NR) limit of the minimal Maxwell CS supergravity
presented here. It has been pointed out that the NR gravities could play an
important role to the understanding of non-relativistic coupled systems in
the boundary. Recently, the NR limit of a three-dimensional Maxwell CS
gravity has been considered in \cite{AFGHZ} where a generalization of the
Extended Bargmann gravity has been obtained.\ Therefore, it would be
interesting to extend the approach of \cite{AFGHZ} to our supergravity case.

\section{Acknowledgment}

This work was supported by the Chilean FONDECYT Projects N$^{\circ }$3170437
(P.C.) and N$^{\circ }$3170438 (E.R.). D.M.P. acknowledges DI-VRIEA for
financial support through Proyecto Postdoctorado 2018 VRIEAPUCV. The authors
would like to thank to L. Ravera for valuable discussions and comments.

\end{document}